\documentclass[preprint,12pt]{elsarticle}
\usepackage{framed,multirow}
\usepackage{array}
\newcolumntype{P}[1]{>{\centering\arraybackslash}p{#1}}
\usepackage{graphicx}
\usepackage{float}
\usepackage{changepage} 
\usepackage{booktabs, multirow}

\usepackage{subcaption,graphicx}
\usepackage{pdfpages}
\usepackage{tikz}
\usepackage{amsmath}
\usepackage[ruled,linesnumbered]{algorithm2e}
\usepackage{amssymb}
\usepackage{latexsym}
\usepackage{mathtools}
\usepackage{hyperref}
\usepackage{diagbox}
\usepackage{makecell}
\usepackage{float}
\hypersetup{
	colorlinks=true,
	linkcolor=blue,
	filecolor=magenta,
	urlcolor=cyan,
}
\usepackage{longtable}
\usepackage{bm}
\usepackage{url}
\usepackage{xcolor}
\usepackage{color} 
\usepackage[dvipsnames]{xcolor}
\definecolor{newcolor}{rgb}{.8,.349,.1}
\usepackage{geometry}

\usepackage{lineno}
\usepackage{caption}
\usepackage{printlen}

\begin{document}

\begin{frontmatter}
\title{Discontinuity-aware physics-informed neural network for phase-field method in three-phase flow with phase change}
\author[1,2]{Guoqiang Lei}
\author[3]{Zhihua Wang}
\author[4]{Lijing Zhou}
\author[1]{D. Exposito}
\author[5,1]{Xuerui Mao\corref{cor1}}
\cortext[cor1]{Corresponding author.}\ead{maoxuerui@sina.com}

\affiliation[1]{
	organization={School of Interdisciplinary Science, Beijing Institute of Technology},
	city={Beijing},
	postcode={100081}, 
	country={China}}
\affiliation[2]{
	organization={School of Mechatronical Engineering, Beijing Institute of Technology},
	city={Beijing},
	postcode={100081}, 
	country={China}}
\affiliation[3]{
	organization={IFP Energies nouvelles},
	addressline={1 et 4 avenue de Bois-Préau},
	city={Rueil-Malmaison},
	postcode={92500},
	country={France}
}
\affiliation[4]{
	organization={School of Aerospace Engineering, Beijing Institute of Technology},
	city={Beijing},
	postcode={100081}, 
	country={China}} 
\affiliation[5]{
	organization={Beijing Institute of Technology (Zhuhai)},
	city={Zhuhai},
	postcode={519088}, 
	country={China}}
\begin{abstract}
Physics-informed neural networks (PINNs) have been applied to simulate multiphase flows, yet they are limited in modeling phase changes and sharp interfaces due to optimization conflicts in the strongly coupled Allen–Cahn, Cahn–Hilliard, and Navier–Stokes equations and the intrinsic smoothness bias of neural representations near discontinuities. To mitigate these limitations, this study presents a discontinuity-aware physics-informed neural network (DPINN) based on the phase-field method to resolve sharp interfaces and phase changes in multiphase flows. It incorporates a discontinuity-aware network architecture to mitigate spectral bias and automatically detect and model sharp interfacial dynamics, and a learnable local artificial viscosity term to stabilize the calculation near steep gradients. During optimization, adaptive time-marching and loss-balancing strategies are employed to reduce long-horizon errors and mitigate gradient conflicts, ensuring accurate capture of phase changes. Numerical experiments on two-phase reversed single-vortex and bubble-rising problems demonstrate that DPINN accurately resolves sharp interfacial dynamics, while conventional PINNs fail to converge. The method is further extended and tested in a three-phase droplet-icing case, where the viscosity and density ratios between ice, water, and air exceed seven and three orders of magnitude, respectively. The predicted phase change dynamics and sharp pointy-tip formation show excellent agreement with the reference, highlighting the robustness of the proposed approach.
\end{abstract}
\end{frontmatter}

\section{Introduction}
\label{sub:1}
Multiphase flows are prevalent in both natural systems and engineering applications \cite{deike2022mass}. A key challenge in the simulation of these flows is to model the interface accurately, whereas the methods can be broadly classified into two branches. The first considers the interface thickness $\xi$ to be zero so that the interface represents a discontinuity, and the physical quantities present a sharp change across the interface, such as the volume-of-fluid \cite{nichols1980sola}, level-set \cite{sussman1994level}, front-tracking \cite{unverdi1992front}, and smoothed particle hydrodynamics methods \cite{maxey2017simulation}. The second, represented by the phase-field method, models the interface as a finite-thickness region governed by partial differential equations (PDEs), such as the Allen–Cahn (AC) equation \cite{cahn1958free} for non-conserved variables and the Cahn–Hilliard (CH) equation \cite{allen1979microscopic} for conserved variables. Contrary to the sharp interface methods, the phase-field method avoids explicit interface tracking and naturally accommodates dynamic boundary conditions on evolving interfaces \cite{biner2017programming}. Since the phase-field model approximates the sharp interface model asymptotically with $O(\xi)$ \cite{huang2020consistent}, the parameter $\xi$ must decrease with the grid size to converge to the desired sharp interface solution, leading to prohibitive computational cost. Furthermore, the steep interfacial gradients lead to the Gibbs phenomenon, resulting in non-physical oscillations near discontinuities \cite{gottlieb1997gibbs}.

Physics-informed neural networks (PINNs) have emerged as a promising alternative framework to simulate multiphase flows. These methods encode physical laws into the loss function, enabling the seamless integration of data and governing equations \cite{raissi2019physics}. In the data-free setting, they can also be trained exclusively using equations and the associated boundary and initial conditions, thus offering a new avenue for surrogate modeling of interfacial dynamics \cite{raissi2024physics}. However, it has been demonstrated that conventional PINNs fail to resolve sharp interfaces \cite{abbasi2025challenges}, whether based on zero-thickness or finite-thickness models \cite{zhou2024self}, due to the inherent smoothness of neural network representations and the optimization paradox near discontinuities, leading to artificially smoothed interfaces or numerical instability \cite{liu2024discontinuity,wang2021understanding}. Numerous methodologies have been developed to address this challenge, broadly categorized into three groups: governing equation reformulations to mitigate gradient explosions at discontinuities \cite{coutinho2023physics,de2024wpinns}, loss and training strategy optimizations to regularize the loss landscape and improve convergence \cite{mao2020physics,ferrer2024gradient}, and architectural enhancements to capture sharp variations \cite{wang2021eigenvector,lei2025discontinuity}. Nevertheless, these improvements remain limited to simplified canonical cases such as the one-dimensional (1D) two-phase Buckley-Leverett problem \cite{abbasi2025challenges}, and their applicability has not been proven in practical multiphase flows, where the intricate interplay of physical forces hinders the resolution of sharp interfaces, typically leading to artificially magnified interface thickness.

Efforts have been devoted to coupling PINNs with the phase-field method, and the early studies primarily focused on improving the accuracy and efficiency of solutions to the strongly nonlinear, high-order, and time-dependent AC or CH equation individually, such as adaptive spatial sampling \cite{wight2020solving}, piecewise time marching \cite{mattey2022novel}, hard constraint \cite{sun2020surrogate}, semi-PINN \cite{zhai2022predicting}, causal training \cite{wang2024respecting}, and high-order optimizer \cite{urban2025unveiling} strategies. For the two-phase coupled CH and Navier-Stokes (NS) system, PINNs have successfully captured the diffuse-interface evolution in 2D bubble dynamics at density ratios up to 1000 \cite{qiu2022physics,zhou2024self} and extended to 3D infinite water tank configurations using axisymmetric coordinates \cite{huang2025phase} or distributed parallel training \cite{qiu2025direct}, complemented by a convergence analysis framework established \cite{buck2025convergence}. However, for three-phase flows governed by the coupled AC–CH formulation, optimization conflicts arising from competing gradient directions hinder network convergence \cite{chen2025pf}. Nevertheless, while optimization strategies such as adaptive loss weighting \cite{chen2025pf} and staggered training \cite{chen2025sharp} have recently been successfully employed to mitigate convergence pathologies in this phase-field formulation, their applicability to the coupled AC–CH–NS system remains unproven, leaving alone the finite-thickness problem inherent in the phase field concept. Consequently, PINNs have not yet been extended to three-phase flow with phase change, such as droplet icing, which is ubiquitous in natural ecosystems and engineering applications (e.g., road traffic, aeronautics, and electrical grids) \cite{wang2025energy,akhtar2023comprehensive}.

To address the aforementioned challenges, including resolving sharp interfaces and phase changes, we develop a discontinuity-aware physics-informed network (DPINN) for the coupled AC–CH–NS system in this work. It incorporates a discontinuity-aware network architecture to mitigate spectral bias and automatically detect and model sharp interfaces, as well as a learnable local artificial viscosity term to stabilize the calculation near steep gradients. During optimization, time-marching and adaptive loss-weighting strategies are employed to reduce long-horizon errors and mitigate gradient conflicts within the coupled system, respectively, ensuring accurate modeling of phase change dynamics.

The remainder of this paper is organized as follows. Section \ref{sub:2} provides an overview of DPINN for phase-field modeling of three-phase flow with phase change. Section \ref{sub:3} demonstrates the superior performances of the proposed method on the two-phase reversed single-vortex and bubble-rising problems compared to conventional PINNs, and extends it to a three-phase droplet-icing case, validated against high-fidelity numerical results. Section \ref{sub:4} concludes this study, discusses its contributions and limitations, and proposes directions for future research.

\section{Methodology}
\label{sub:2}
\begin{figure}[!h]
	\centering
	\includegraphics[width=\linewidth]{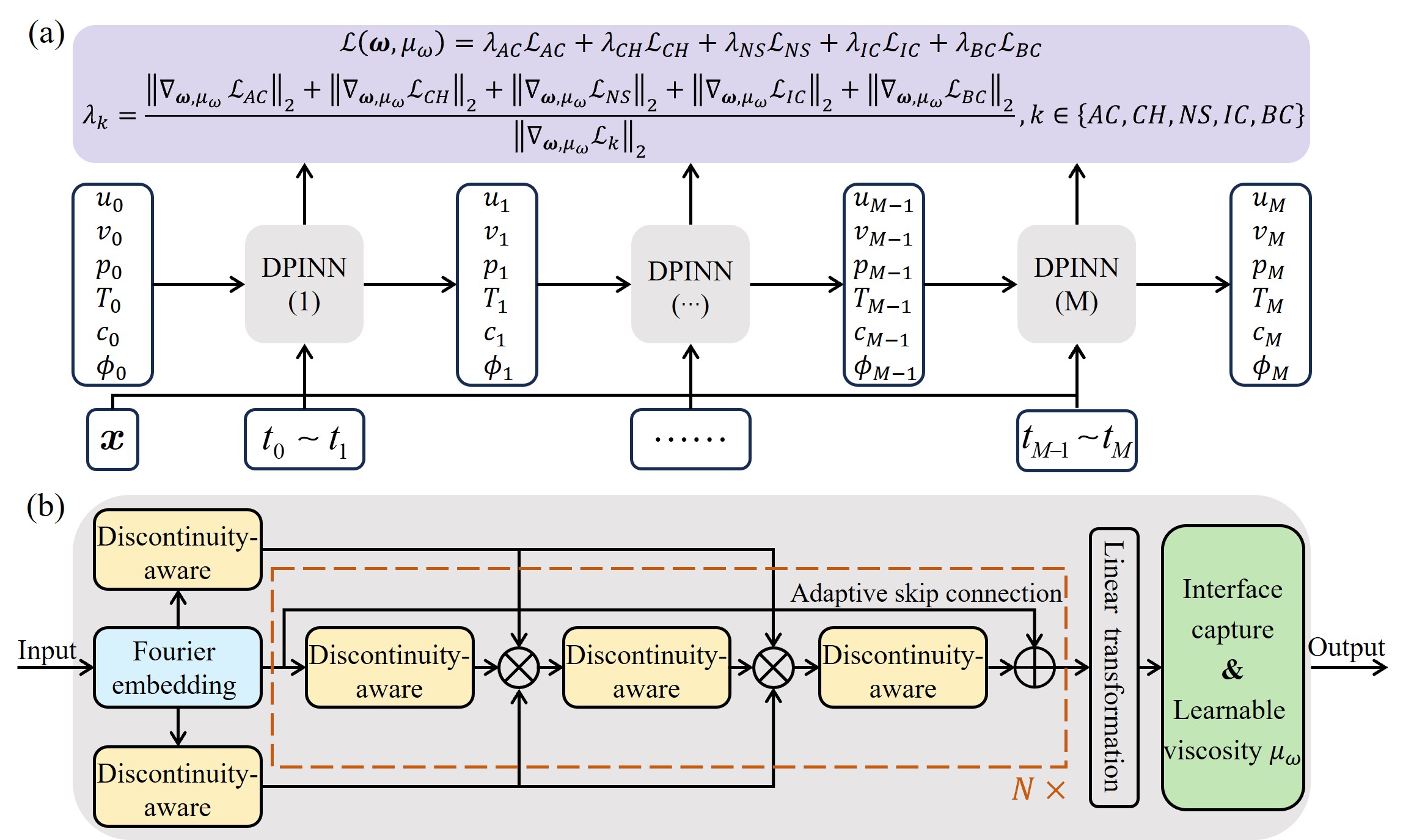}
	\caption{(a) Flowchart of DPINN applied to a 2D unsteady three-phase flow with phase change. (b) Model architecture of DPINN.}
	\label{fig: flowchat}
\end{figure}

As mentioned above, the proposed DPINN method serves as an artificial intelligence (AI)-based multiphase flow simulator, capable of accurately resolving sharp interfaces and phase changes through enhanced network architecture and optimization strategies. Figure \ref{fig: flowchat}a presents the flowchart of the proposed DPINN method for phase-field modeling of three-phase flows governed by the coupled AC–CH–NS equations, without relying on any data-driven prior information (Section \ref{sub:2.1}). In terms of network architecture (Figure \ref{fig: flowchat}b), we introduce Fourier embedding and discontinuity-aware layers to automatically detect and approximate sharp interfaces (Section \ref{sub:2.2}), and a learnable artificial viscosity limited to interfacial regions to improve solution stability (Section \ref{sub:2.2.5}). In terms of optimization, time-marching and adaptive loss-weighting strategies are employed to mitigate long-horizon errors and gradient conflicts among the coupled equations, respectively, for the accurate modeling of phase change dynamics (Section \ref{sub:2.3}).

\subsection{PINN-based phase-field model for three-phase flow with phase change}
\label{sub:2.1}
Considering the ice-water-air three-phase flow, the phase-field method and its PINN-based form are briefly reviewed in this section. Phase-field variables are introduced to represent the continuous variation of properties across the interface governed by the non-conserved AC and conserved CH equations, whilst the fluid dynamics is modeled by the NS equations consisting of the continuity, momentum, and energy equations.

The AC equation \cite{allen1979microscopic} is used to model the phase change from water to ice:
\begin{equation}
	\frac{\partial c}{\partial t}+\bm{u}\cdot \nabla c={M_c}\left[6\sigma_c\xi_c{\nabla^2}c-\frac{3\sigma_c}{\xi_c}{G'_1}(c)-\frac{T_M-T}{T_M}H{G'_2}(c) \right],
	\label{eq: AC}
\end{equation}
where the phase-field order parameter $c$ is employed to distinguish water and air $(c=0)$ from ice $(c=-1)$, $t$ is time, $\bm{u}$ is the velocity vector, $M_c$, $\sigma_c$, and $\xi_c$ represent the mobility, surface tension coefficient, and thickness of the ice-water interface, respectively, $H$ denotes the latent heat per unit volume, $T$ is the temperature field, and $T_M$ is the melting point. The double-well function ${G_1}(c)={(c+1)^2}{c^2}$ describes the bulk free energy density, and ${G_2}'(c)=30{G_1}(c)$, where the prime denotes the derivative \cite{zhang2022phase}.

The CH equation \cite{cahn1958free} is employed to model the evolution of the interface between water and air:
\begin{equation}
	\begin{aligned}
	&\frac{\partial \phi }{\partial t}+\bm{u}\cdot \nabla \phi ={M_\phi}{\nabla^2}{u_\phi}, \\
	&{u_\phi}=-\frac{3 \sigma_\phi}{2\sqrt{2}{\xi_\phi }}\left[\xi_\phi^2 {\nabla^2}\phi -G'_3(\phi) \right],
	\end{aligned}
	\label{eq: CH}
\end{equation}
where $\phi$ is the other phase-field order parameter that distinguishes ice and water ($\phi=1$) from air ($\phi=-1$), ${M_\phi}$, $\sigma_\phi$, and $\xi_\phi$ represent the mobility, surface tension coefficient, and thickness of the water-air interface, respectively, ${u_\phi}$ is the chemical potential, and $G_3(\phi)=\frac{1}{4}{(\phi-1)^2}{(\phi+1)^2}$ is the double-well function \cite{wang2025energy}.

The volume fractions of the air, water, and ice in the proposed model are defined as $\alpha_{air}=(1-\phi)/2$, $\alpha_{water}=(1+c)(1+\phi)/2$, $\alpha_{ice}=(-c)(1+\phi)/2$, respectively, and $\alpha_{air} + \alpha_{water} + \alpha_{ice} = 1$. The density $\rho$, viscosity $\mu$, thermal conductivity $k$, and specific heat capacity $c_p$ are obtained by linear interpolation of the phase volume fractions \cite{huang2022consistent}:
\begin{equation}
	\begin{aligned}
		\mathcal{P}(c,\phi)=\alpha_{air} \mathcal{P}_{air}+\alpha_{water} \mathcal{P}_{water}+\alpha_{ice} \mathcal{P}_{ice},
		\quad
		\mathcal{P}\in\{\rho,\mu,k,c_p\}.
	\end{aligned}
	\label{eq: rho}
\end{equation}

The continuity equation governing all the phases is
\begin{equation}
	\frac{\partial \rho}{\partial t}+\nabla \cdot (\rho \bm{u})=0.
	\label{eq: mass}
\end{equation}

The momentum equation in a conservation form is
\begin{equation}
	\frac{\partial(\rho \bm{u})}{\partial t}+\nabla \cdot(\rho \bm{u} \bm{u}^\top)=-\nabla p+\nabla \cdot\left\{\mu\left[\nabla \boldsymbol{u}+(\nabla \bm{u})^\top-\frac{2}{3}\nabla \cdot \bm{u} \bm{I}\right]\right\}-\varepsilon_\phi \nabla^2 \phi \nabla \phi-\varepsilon_c \nabla^2 c \nabla c+\boldsymbol{F}_b,
	\label{eq: momentum}
\end{equation}
where $\varepsilon_\phi=\frac{3}{2\sqrt{2}}\sigma_\phi\xi_\phi$ and $\varepsilon_c=\frac{3}{2\sqrt{2}}\sigma_c \xi_c$ are the mixing energy density parameters of the water-air interface and the water-ice interface, respectively, $\bm{I}$ is the unit diagonal tensor, and $\bm{F}_b=\rho \bm{g}$ is the body force, with $\bm{g}$ denoting the gravitational acceleration.

The energy equation is employed in this paper to guarantee the energy conservation of the system \cite{wang2025energy}:
\begin{equation}
	\frac{\partial\left(\rho c_p T\right)}{\partial t}+\nabla \cdot\left(\rho c_p T \bm{u}\right)=\frac{\partial \alpha_{ice}}{\partial t} H+\nabla \cdot(\kappa \nabla T)+\Phi+\bm{u} \cdot \bm{F}_b,
	\label{eq: energy}
\end{equation}
where the first term on the right side depicts the latent heat due to phase change in the icing process. $\Phi =\mu\left[ \nabla \bm{u}+(\nabla \bm{u})^\top-2/3 \nabla \cdot \bm{u}\bm{I}\right] ):\nabla \bm{u}$ is the viscous dissipation rate.

In the PINN framework, the AC (Eq. \ref{eq: AC}), CH (Eq. \ref{eq: CH}), and NS (Eqs. \ref{eq: mass}-\ref{eq: energy}) equations are formulated as equality constraint operators, denoted by $\mathcal{N}_{AC}$, $\mathcal{N}_{CH}$, and $\mathcal{N}_{NS}$, respectively. These operators are incorporated into the loss function to enforce compliance with the governing equations and associated constraints. In contrast to numerical methods that solve PDEs iteratively, PINNs take the spatial coordinate vector $\bm{x} \in \Omega$ and time $t \in[0, \mathcal{T}]$ as input and optimizes the neural network parameters $\bm{w}$ to minimize the loss function $\mathcal{L}(\bm{w})$, resulting in the optimal parameters $\bm{w}^*$ and approximate solutions:
\begin{equation}
	\begin{aligned}
		\bm{w}^*=\underset{\bm{w}}{\operatorname{argmin}} \mathcal{L}(\bm{w}) &= \underset{\bm{w}}{\operatorname{argmin}} \left( \lambda_{AC} \mathcal{L}_{AC} + \lambda_{CH} \mathcal{L}_{CH} + \lambda_{NS} \mathcal{L}_{NS} + \lambda_{IC} \mathcal{L}_{IC} + \lambda_{BC} \mathcal{L}_{BC} \right), \\
		\mathcal{L}_{k} &= \| \mathcal{N}_{k}\left[\bm{u}(\bm{x}, t; \bm{w}) \right] \|_2, \,\; \bm{x} \in \Omega, t \in(0, \mathcal{T}], k\in\{AC,CH,NS\}, \\
	\end{aligned}
	\label{eq: loss}
\end{equation}
where $\Omega$ represents the computational domain, $\mathcal{T}$ is the final time, and $\left\|\cdot\right\|_2$ denotes the $L_2$ norm; $\mathcal{L}_{AC}$, $\mathcal{L}_{CH}$, $\mathcal{L}_{NS}$, $\mathcal{L}_{IC}$, and $\mathcal{L}_{BC}$ correspond to the loss components for the AC, CH, and NS equations, and the initial and boundary conditions, respectively; and the non-negative weights $\lambda_{AC}$, $\lambda_{CH}$, $\lambda_{NS}$, $\lambda_{IC}$, and $\lambda_{BC}$ control the trade-off among these components. Provided the solution is unique, the output of the neural network converges to the solution of the PDE system \cite{raissi2019physics}.
\subsection{Discontinuity-aware physics-informed neural network}
\label{sub:2.2}
Conventional PINNs are unsuitable for multiphase flows with sharp interfaces, where interfacial discontinuities lead to an optimization paradox and induce oscillatory loss during training \cite{wang2021understanding,wang2022and,liu2024discontinuity}. To address this issue, we propose DPINN to automatically detect and model sharp spatial transitions and rapid temporal evolution, based on a residual-adaptive architecture \cite{wang2024piratenets} specifically designed to facilitate stability and convergence in deeper neural network architectures, and an additional learnable local artificial viscosity term that introduces dissipation to stabilize solutions near steep gradients.

Figure \ref{fig: flowchat}b presents the architecture of DPINN. Due to spectral bias \cite{rahaman2019spectral}, standard neural networks tend to prioritize low-frequency components, failing to resolve high-frequency ones that construct discontinuities. Therefore, first, the network input $\bm{\chi}_{input}=(\bm{x},t)$ is encoded into a high-dimensional feature space through a Fourier embedding layer with learnable sampled frequencies to enhance sensitivity to discontinuous features. This embedding has proven to be effective in mitigating the optimization paradox near steep gradients \cite{tancik2020fourier}, and is defined as:

\begin{equation}
	{\mathcal{F}}(\bm{\chi}_{input})={\frac{1}{\sqrt{m}}}\left[\sin \left(2 \pi \bm{W}_f \bm{\chi}_{input}\right), \cos \left(2 \pi \bm{W}_f \bm{\chi}_{input}\right)\right]^\top \in \mathbb{R}^{2m},
\end{equation}
where $\bm{W}_f \in \mathbb{R}^{m \times d_{input}}$ is the learnable frequency matrix, initially sampled from a zero-mean, unit-variance Gaussian distribution, and $m$ and $d_{input}$ denote the number of frequency components and the input dimension, respectively.

Then, ${\mathcal{F}}(\bm{\chi}_{input})$ is mapped into two discontinuity-aware layers that act as gates in each residual block to enhance the trainability and improve the convergence of DPINN \cite{wang2021understanding}. These layers are defined as:
\begin{equation}
	\begin{aligned}
		\bm{U}= & \operatorname{DyT}\left(\bm{W}_U \mathcal{F}(\bm{\chi}_{input})+\bm{b}_U\right), \\
		\bm{V}= & \operatorname{DyT}\left(\bm{W}_V \mathcal{F}(\bm{\chi}_{input})+\bm{b}_V\right), \\
	\end{aligned}
	\label{eq: UV}
\end{equation}
where $\bm{W}_U$, $\bm{W}_V \in \mathbb{R}^{2m \times 2m}$ and $\bm{b}_U$, $\bm{b}_V \in \mathbb{R}^{2m}$ denote the linear weight matrices and bias vectors, respectively, with subscripts indexing their corresponding networks. The learnable Dynamic Tanh activation function ($\operatorname{DyT}$) is capable of sensing and modeling discontinuity features by dynamically adjusting the input activation range \cite{zhu2025transformers}, defined as:
\begin{equation}
	\operatorname{DyT}(\bm{\chi})={\bm{w}_t}\tanh(\bm{w}_a \bm{\chi})+\bm{w}_b,
	\label{eq: DYT}
\end{equation}
where $\bm{w}_t$, $\bm{w}_a$, $\bm{w}_b \in \mathbb{R}^{2m}$ are the trainable parameters.

Let $\bm{\chi}^1={\mathcal{F}}(\bm{\chi}_{input})$ be the input to the first discontinuity-aware residual-adaptive network block. The network consists of $N$ such blocks, where the $n$-th $(n=1,\ldots,N)$ block maps the input $\boldsymbol{\chi}^n$ to the output $\boldsymbol{\chi}^{n+1}$ according to:
\begin{equation}
	\begin{aligned}
		\bm{f}_1^n &= \operatorname{DyT}\!\left(\bm{W}_1^n \bm{\chi}^n+\bm{b}_1^n\right),&\quad
		&\bm{z}_1^n = \bm{f}_1^n \odot \bm{U} + \left(1-\bm{f}_1^n\right) \odot \bm{V},
		\\
		\bm{f}_2^n &= \operatorname{DyT}\!\left(\bm{W}_2^n \bm{z}_1^n+\bm{b}_2^n\right),&\quad
		&\bm{z}_2^n = \bm{f}_2^n \odot \bm{U} + \left(1-\bm{f}_2^n\right) \odot \bm{V},
		\\
		\bm{f}_3^n &= \operatorname{DyT}\!\left(\bm{W}_3^n \bm{z}_2^n+\bm{b}_3^n\right),&\quad
		&\bm{\chi}^{n+1} = w_\alpha^n \bm{f}_3^n + \left(1-w_\alpha^n\right) \bm{\chi}^n,
	\end{aligned}
	\label{eq: Block}
\end{equation}
where $\odot$ denotes a point-wise multiplication, the superscript indexes the block, the numeric subscript indexes different networks within the same block, and $w_\alpha^n \in \mathbb{R}$ is a trainable scalar parameter that modulates the nonlinearity of the $n$-th block by acting as an adaptive skip-connection weight. When $w_\alpha^n=0$, $\bm{\chi}^{n+1}=\bm{\chi}^n$ is an identity map, whereas $w_\alpha^n=1$ yields a fully nonlinear mapping without shortcuts. The final network output is defined by a linear transformation of the $N$-th discontinuity-aware residual-adaptive block output $\bm{\chi}^{N+1}$:
\begin{equation}
	\bm{\chi}_{output}= \bm{W}_{output}\bm{\chi}^{N+1}, \\
	\label{eq: output}
\end{equation}
where $\bm{W}_{output} \in \mathbb{R}^{2m \times d_{output}}$ is the linear weight matrix and $d_{output}$ denotes the output dimension.

\subsection{Learnable local artificial viscosity}
\label{sub:2.2.5}
Artificial viscosity is a classical approach in computational fluid dynamics for improving numerical stability near discontinuities \cite{godunov1959finite}. It has also been integrated into PINNs as a physics-based modification strategy to approximate sharp interfacial dynamics by explicitly introducing dissipation to smooth sharp spatial transitions and rapid temporal evolution \cite{abbasi2025challenges}. Specifically, an additional viscosity term in the governing system (Eqs. \ref{eq: AC}, \ref{eq: CH}, \ref{eq: mass}-\ref{eq: energy}) can be reformulated as $\nabla \cdot (\mu_w \nabla \bm{Q})$, where $\mu_w \ge 0$ is the network-predicted artificial viscosity distribution, and $\bm{Q}=[\rho,\rho u, \rho v, \rho c_p T, c, \phi]^\top$ denotes the conservative state vector for the three-phase flow with phase change. However, selecting an appropriate magnitude and activation region for $\mu_w$ is challenging: an excessively large value induces over-dissipation and reduces accuracy, whereas an insufficient one fails to mitigate numerical instability or converge to the sharp interface solution. Theoretically, $\mu_w$ should be limited to the interface and negligible in smooth regions \cite{lei2025discontinuity}.

In this study, benefiting from the unique advantages of the phase-field method, the region of application is readily identified through a thresholding procedure based on the phase-field order parameter, while its magnitude is predicted by the network rather than directly inferring the entire distribution \cite{coutinho2023physics,wassing2024physics}, as defined by:
\begin{equation}
	\mu_w=w_\mu [(-1+\delta \leq c \leq-\delta) \vee(-1+2 \delta \leq \phi \leq 1-2 \delta)],
	\label{eq: mu}
\end{equation}
where $w_\mu \ge 0$ is a learnable network parameter that controls the dissipation magnitude, and the latter term serves as an interface sensor limiting the active region of the artificial viscosity, with the threshold hyperparameter set to $\delta=0.25$.
\subsection{Adaptive time-marching and loss-balancing optimization strategies}
\label{sub:2.3}
In the simulation of unsteady three-phase flows with phase change, PINNs suffer from long-horizon errors \cite{chen2024pinn} and gradient conflicts from the unbalanced loss in the strongly coupled AC-CH-NS system \cite{wang2021understanding,yu2020gradient}. To mitigate these issues, the adaptive time-marching scheme \cite{mattey2022novel} is employed as a piecewise manifold approximation method to reduce long-term error accumulation, while the loss-balancing optimization strategy \cite{mcclenny2023self,bischof2025multi} addresses competing gradient directions in the coupled system.

In the adaptive time-marching strategy, the temporal domain $\mathcal{T}$ is divided into $M$ consecutive subdomains $(\mathcal{T}_1=[t_0,t_1], \cdots, \mathcal{T}_M=[t_{M-1},t_M])$, each associated with an identical DPINN and trained sequentially within its assigned interval, where the predicted solution at the end of one segment is used as the initial condition to fine-tune the network in the next until the entire temporal domain is covered, as illustrated in Figure \ref{fig: flowchat}a. The individual networks trained in different time segments predict the solutions in their respective domains, collectively forming the complete solution of the coupled PDE system, which improves the overall accuracy compared with using a single network, albeit at the expense of higher training cost \cite{qiu2022physics,zhou2024self}.

In multi-objective optimization of the coupled AC-CH-NS system, manually tuning weighting coefficients becomes intractable as the number of objectives increases \cite{bischof2025multi}, while the strongly coupled AC-CH-NS equations exhibit pathological convergence due to conflicting gradient directions \cite{chen2025pf,chen2025sharp}. To mitigate these issues, the self-adaptive loss-balancing strategy based on gradient statistics is employed to dynamically adjust loss weights, minimizing manual intervention and maintaining training balance for multiphase flow with phase change. These weights are defined by:

\begin{equation}
	\lambda_k=\frac{{\left\| {\nabla_{\bm{\omega},\mu_w }}{\mathcal{L}_{AC}} \right\|}_2+{\left\| {\nabla_{\bm{\omega},\mu_w }}{\mathcal{L}_{CH}} \right\|}_2+{\left\| {\nabla_{\bm{\omega},\mu_w }}{\mathcal{L}_{NS}} \right\|}_2+{\left\| {\nabla_{\bm{\omega},\mu_w }}{\mathcal{L}_{IC}} \right\|}_2+{\left\| {\nabla_{\bm{\omega},\mu_w }}{\mathcal{L}_{BC}} \right\|}_2}{{\left\| {\nabla_{\bm{\omega},\mu_w }}{\mathcal{L}_{k}} \right\|}_2},
	\label{eq: lambda_k}
\end{equation}
where $\bm{\omega}$ is the network parameters, $\mu_w$ represents the learnable viscosity distribution, and $\mathcal{L}_{k}$ ($k \in \left\{ AC, CH, NS, IC, BC \right\}$) denote the loss terms associated with the AC, CH, and NS equations, and the initial and boundary conditions, respectively. With this method, the weighting coefficients $\lambda_k$ are adaptively tuned to equalize the gradient norms across all constituent loss terms, preventing bias toward any particular term during training, such that:
\begin{equation}
	\resizebox{\hsize}{!}{%
		$\| \lambda_{AC} \nabla_{\omega,\mu_w} \mathcal{L}_{AC} \|_2 \!=\! \| \lambda_{CH} \nabla_{\omega,\mu_w} \mathcal{L}_{CH} \|_2 \!=\! \| \lambda_{NS} \nabla_{\omega,\mu_w} \mathcal{L}_{NS} \|_2 \!=\! \| \lambda_{IC} \nabla_{\omega,\mu_w} \mathcal{L}_{IC} \|_2 \!=\! \| \lambda_{BC} \nabla_{\omega,\mu_w} \mathcal{L}_{BC} \|_2$.
	}
	\label{eq:weight}
\end{equation}
These weighting coefficients are updated at user-specified intervals (typically every 100 iterations) to stabilize stochastic gradient descent while limiting computational overhead.
\section{Results}
\label{sub:3}
This section evaluates the performance of the proposed method in solving the coupled AC–CH–NS equations for multiphase flows with sharp interfaces and phase changes. As a preliminary assessment, the method is applied to two two-phase cases: a reversed single vortex and a classical bubble-rising case, to quantify its accuracy in capturing sharp interfacial dynamics and robustness under large interfacial deformations, compared to conventional PINNs. Finally, the method is applied to a realistic three-phase droplet-icing case with viscosity and density ratios over seven and three orders of magnitude, respectively. 

Due to sharp transitions, three error metrics are employed to measure the accuracy of the solution, defined as: 
\begin{equation}
	\begin{aligned}
		R_1=\frac{\left\| u_{true}-u \right\|_1}{\left\| u_{true} \right\|_1}, \quad R_2=\frac{\left\| u_{true}-u \right\|_2}{\left\| u_{true} \right\|_2}, \quad R_{\infty}=\frac{\left\| u_{true}-u \right\|_\infty}{\left\| u_{true} \right\|_\infty},
	\end{aligned}
	\label{eq: three metrics}
\end{equation}
where $\left\|\cdot \right\|_1$, $\left\|\cdot \right\|_2$, and $\left\|\cdot \right\|_\infty$ represent the $L_1$, $L_2$, and $L_\infty$ norms, respectively.

\subsection{Two-phase reversed single vortex}
\label{sub:3.1}
A reversed single vortex case is used to evaluate the performance of DPINN in resolving interface dynamics governed by the CH equation (Eq. \ref{eq: CH}). The velocity field is given by the stream function:

\begin{equation}
	\Psi (x,y,t)=\frac{1}{\pi } \sin^2\left( \pi x\right) \, \sin^2\left( \pi y\right) \, \cos\left( \frac{\pi t}{\mathcal{T}_0}\right),
	\label{eq:vortex_uv}
\end{equation}
where the rotational period $\mathcal{T}_0=2$ s. The initial circular interface is stretched by the given flow field, reaches maximum deformation at $t=\mathcal{T}_0/2$, and reverses to its initial position and shape at $t=\mathcal{T}_0$, maintaining a constant area in the whole process. The velocities in the $x$- and $y$-directions are defined as $u(x,y,t)=\partial \Psi/ \partial y$ and $v(x,y,t)=-\partial \Psi/ \partial x$, respectively. The spatial domain $\Omega = [0,1]$ m$\times$ $[0,1]$ m is discretized with 200 uniform grid points in each spatial direction, while the temporal domain $\mathcal{T} = [0,1]$ s is treated as a single temporal subdomain $M=1$ covering half the oscillation period with 100 uniform time steps, with periodic boundary conditions applied in both $x$- and $y$-directions. Initially, a circular interface is centered at $(x_0, y_0)=(0.5,0.75)$ with a radius $r_0=0.15$ m. The initial phase-field distribution is defined as $\phi=-1$ inside the circle (air phase) and $\phi=1$ outside (water phase):

\begin{equation}
	{\phi}(x,y,t=0) = \begin{cases}
		{-1} & \text{if } (x - x_0)^2 + (y - y_0)^2 \le r_0^2,\\ 
		{1} & \text{otherwise}.
	\end{cases}
	\label{eq: vortex_init}
\end{equation}

In this case, the network architecture comprises a Fourier embedding with 64 learnable frequencies, $N=1$ discontinuity-aware adaptive residual block (containing 3 discontinuity-aware layers with 128 neurons each), a learnable artificial-viscosity coefficient, and a linear output layer with a single node corresponding to the phase-field $\phi$. Training is performed for 20000 iterations using the AdamW optimizer under a OneCycleLR schedule, where the learning rate rises from $10^{-5}$ during a 10\% warm-up to a peak of $10^{-3}$ and then decays back to $10^{-5}$, on a Hygon G86 7285 with 32 GB of memory.

\begin{figure}[!h]
	\centering
	\includegraphics{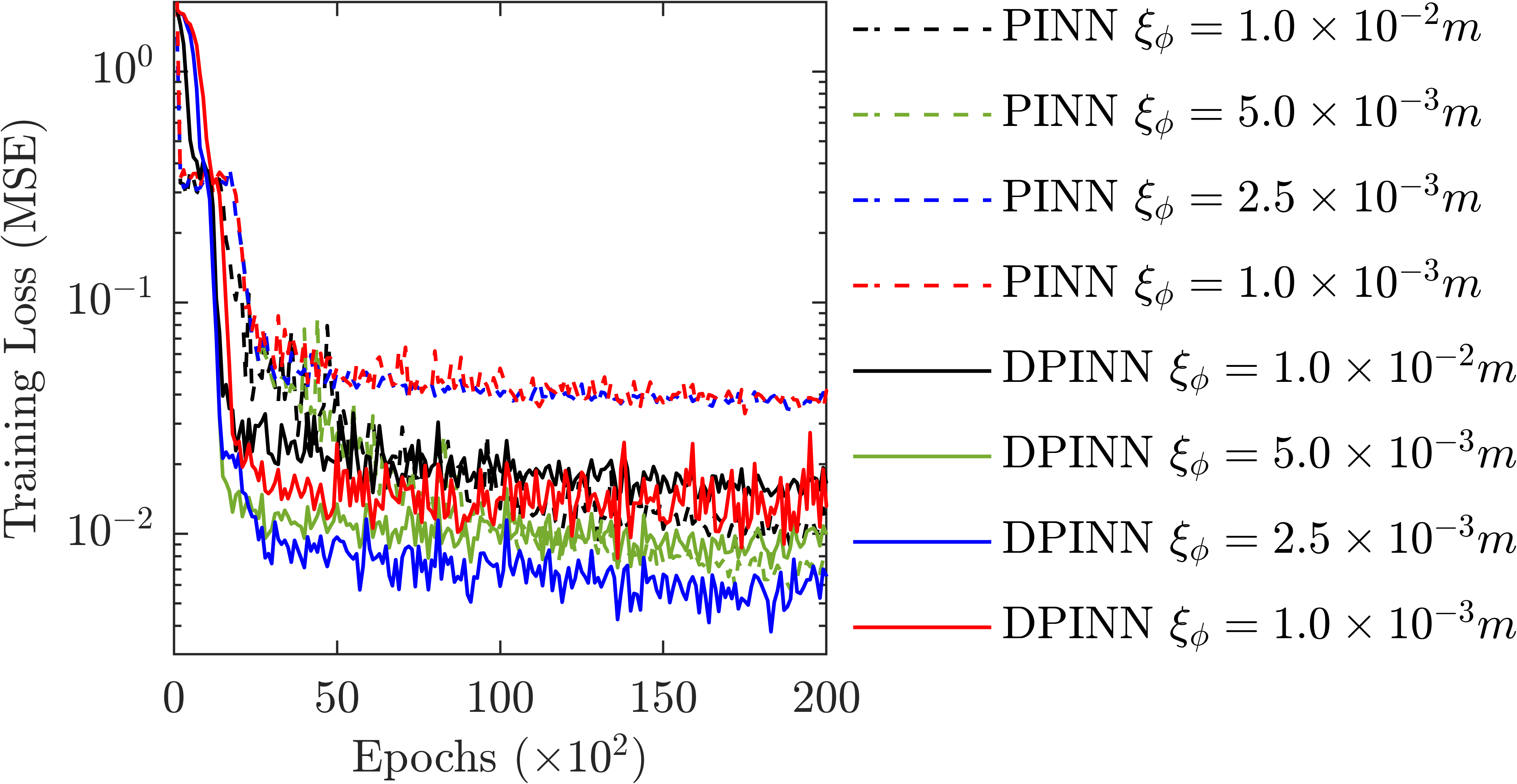}
	\caption{Evolution of the loss function with training epochs.}
	\label{fig: vortex_train_loss}
\end{figure}

The evolution of the training loss and solution of the conventional PINN and the proposed DPINN are examined for four interface thicknesses ($1.0 \times 10^{-2}$ m, $5.0 \times 10^{-3}$ m, $2.5 \times 10^{-3}$ m, and $1.0 \times 10^{-3}$ m), representing a transition from smooth to sharp interfaces, under a uniform mobility number of $M_\phi = 10^{-4}$ $m^4 \cdot N^{-1}\cdot s^{-1}$. As shown in Figure \ref{fig: vortex_train_loss}, for $\xi_\phi = 1.0 \times 10^{-2}$ m and $5.0 \times 10^{-3}$ m, where interfacial dissipation is significant, both PINN and DPINN exhibit comparable convergence behavior. However, as the interface thickness further decreases to $\xi_\phi = 2.5 \times 10^{-3}$ m and $1.0 \times10^{-3}$ m, the conventional PINN fails to converge to the levels achieved with thicker interfaces due to the optimization paradox near sharp interfaces, whereas DPINN achieves a loss approximately one order of magnitude lower than that of the conventional PINN.

\begin{figure}[!h]
	\centering
	\includegraphics{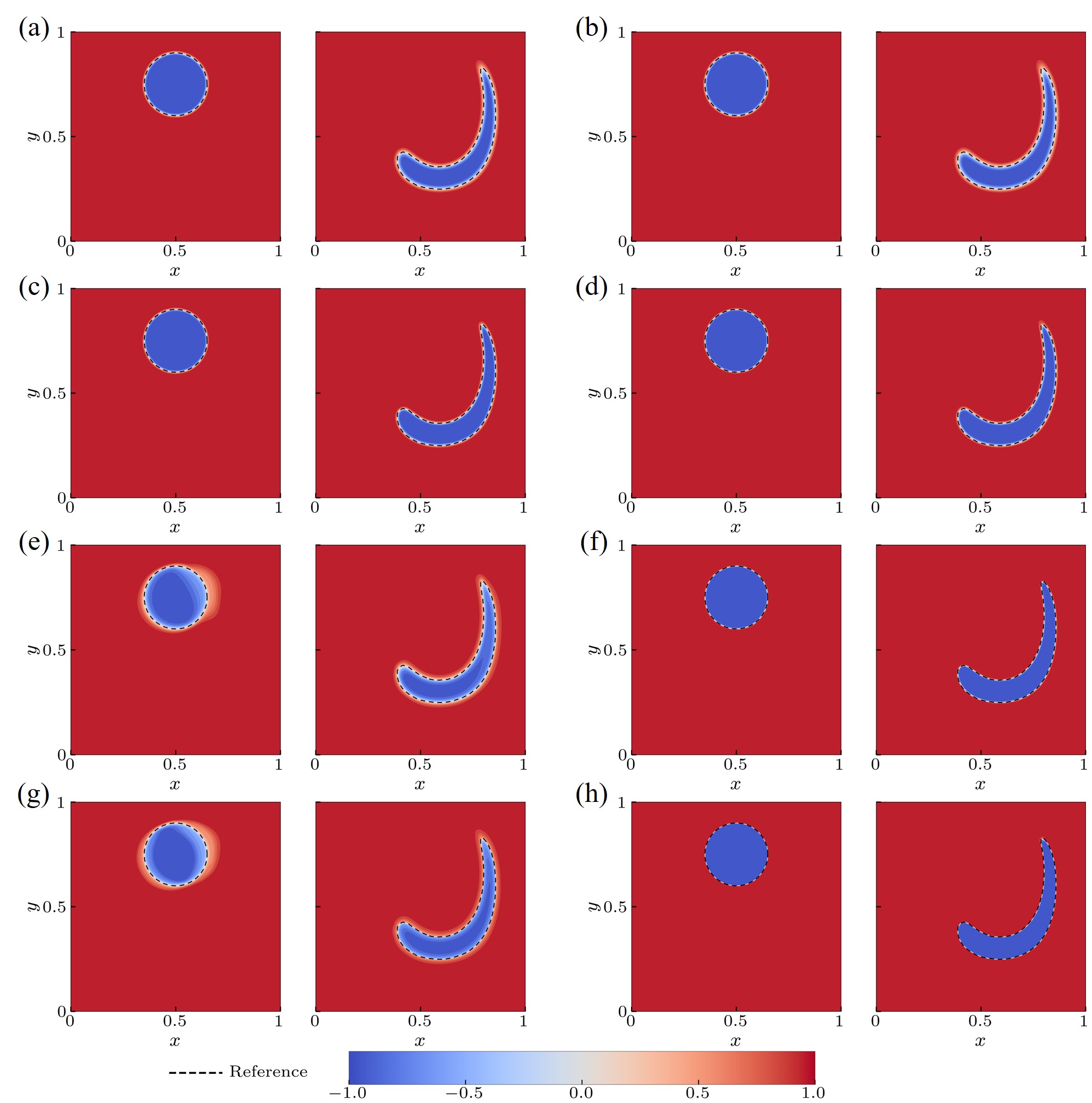}
	\caption{Predicted results of phase-field variable $\phi$ at $t=0$ s (left) and $1$ s (right). (a), (c), (e), and (g) present the PINN solutions for $\xi_\phi=0.01$ m, $0.005$ m, $0.0025$ m, and $0.001$ m, respectively, while (b), (d), (f), and (h) show the corresponding DPINN solutions. Dotted lines represent the reference numerical results from Ref.~\cite{huang2020consistent}.}
	\label{fig: vortex_conrourf}
\end{figure}

Figure \ref{fig: vortex_conrourf} compares PINN and DPINN results for different interface-thickness parameters $\xi_\phi$ at $t=0$ s and $t=1$ s against reference numerical results from Ref.~\cite{huang2020consistent}. For $\xi_\phi =1.0 \times 10^{-2}$ m and $5.0 \times 10^{-3}$ m, both the conventional PINN and DPINN accurately capture the temporal evolution of the diffuse interface, with errors gradually decreasing as the interface thickness decreases. However, for $\xi_\phi = 2.5 \times 10^{-3}$ m and $1.0 \times 10^{-3}$ m, the conventional PINN fails to converge to the sharp interface solution, causing local interfacial errors to propagate throughout the entire computational domain, whereas DPINN accurately resolves these sharp interface dynamics, with errors steadily decreasing. 

Table \ref{tab:error_norms_phi} reports the $R_1$, $R_2$, and $R_\infty$ area-error norms over the entire time horizon relative to the theoretically constant area of the reversed single vortex. As $\xi_\phi$ decreases from a thick interface to a sharp one, PINN errors initially decrease but subsequently increase due to convergence issues in steep interfacial regions, whereas DPINN errors decrease consistently.

\begin{table}[!h]
	\centering
	\renewcommand\arraystretch{0.8}
	\begin{tabular}{@{}lcccc@{}}
		\toprule
		Model & $\xi_\phi$/m & $R_1$ & $R_2$ & $R_{\infty}$ \\
		\midrule
		\multirow{4}{*}{\centering PINN}
		& 0.0100 & 0.0192 & 0.0197 & 0.0260 \\
		& 0.0050 & 0.0123 & 0.0125 & 0.0155 \\
		& 0.0025 & 0.0282 & 0.0286 & 0.0339 \\
		& 0.0010 & 0.0289 & 0.0302 & 0.0374 \\
		\midrule
		\multirow{4}{*}{\centering DPINN} 
		& 0.0100 & 0.0161 & 0.0166 & 0.0216 \\
		& 0.0050 & 0.0091 & 0.0094 & 0.0128 \\
		& 0.0025 & 0.0043 & 0.0045 & 0.0064 \\
		& 0.0010 & 0.0043 & 0.0044 & 0.0055 \\
		\bottomrule
	\end{tabular}
	\caption{Area-error norms of the predicted phase-field at different interface-thickness parameters $\xi_\phi$ for PINN and DPINN.}
	\label{tab:error_norms_phi}
\end{table}

\subsection{Two-phase bubble rising}
\label{sub:3.2}
The bubble rising problem serves as a challenging case for accurate simulations of multiphase flows with large interfacial deformations. In this case, an air bubble released inside a closed water tank rises due to buoyancy and continuously deforms under the competing influences of inertia, viscous stresses, surface tension, and gravity, involving large density and viscosity ratios. The governing equations are the coupled CH-NS equations (Eqs. \ref{eq: CH}, \ref{eq: mass}, and \ref{eq: momentum}). The spatial domain is defined as $\Omega = [0,1]$ m$\times$ $[0,2]$ m, discretized into 200 and 400 grid points in the $x$- and $y$-directions, respectively, and the temporal domain $\mathcal{T}=[0,3]$ s is divided into $M=2$ subdomains, $\mathcal{T}_1=[0,1.5]$ s and $\mathcal{T}_2=[1.5,3]$ s, each uniformly discretized with 100 time steps. No-slip boundary conditions are imposed on the top and bottom walls, while free-slip conditions are applied on the left and right boundaries. Initially, an air bubble with a radius of $r_0=0.25$ m is placed at $(x_0,y_0)=(0.5,0.5)$ in water with zero velocity, as defined in Eq. \ref{eq: vortex_init}. 

\begin{figure}[!h]
	\centering
	\includegraphics{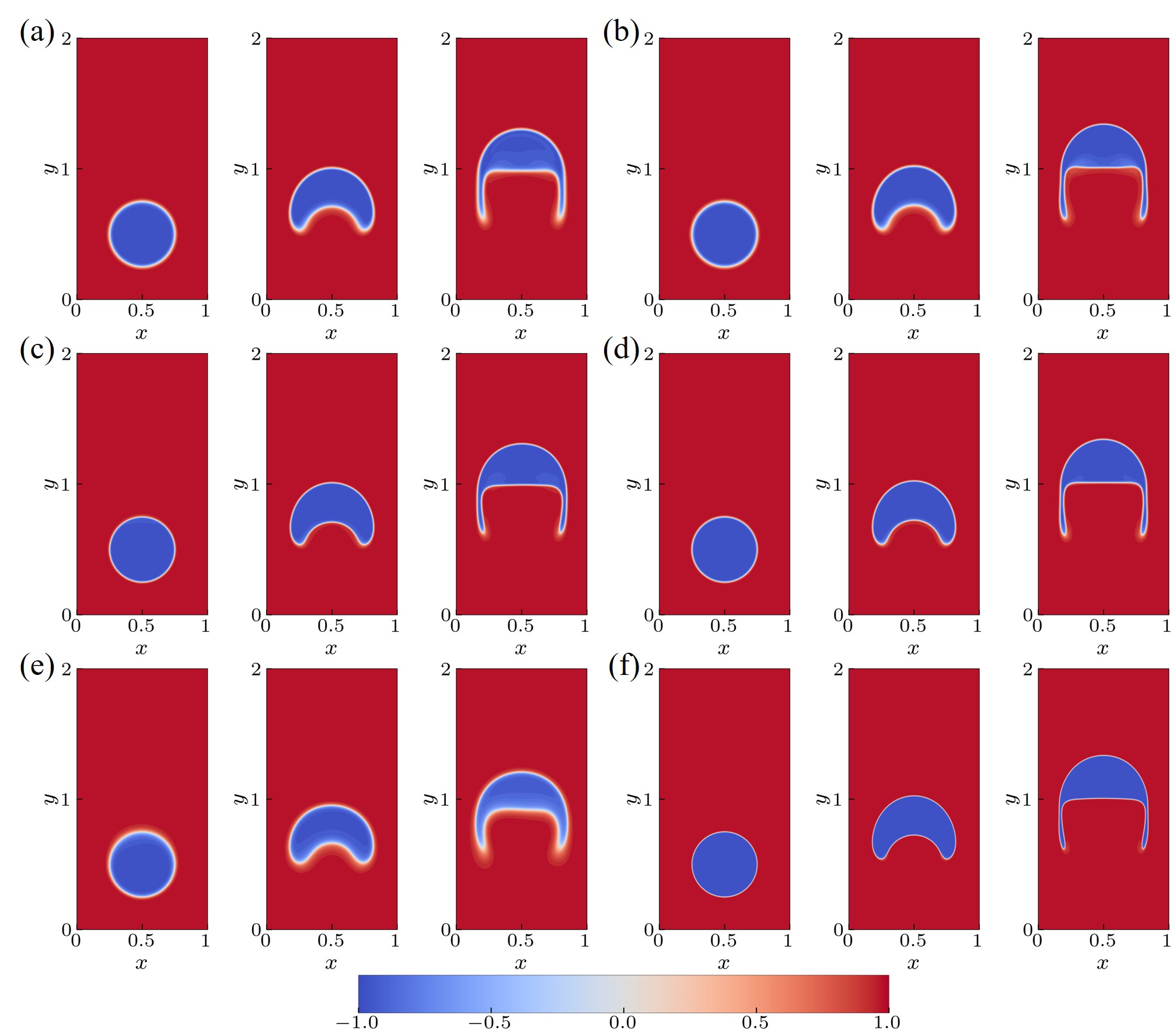}
	\caption{Predicted results of phase-field variable $\phi$ at $t=0$ s, $1.5$ s, and $3$ s (left to right). (a), (c), and (e) show the PINN solutions for $\xi_\phi=0.01$ m, $0.005$ m, and $0.0025$ m, respectively, while (b), (d), and (f) show the corresponding DPINN solutions.}
	\label{fig: bubble_conrourf}
\end{figure}

Following the benchmark case of 2D bubble dynamics \cite{aland2012benchmark}, the density and viscosity of air inside the bubble are set to $\rho_{air} = 1\,\mathrm{kg\cdot m^{-3}}$ and $\mu_{air} = 0.1\,\mathrm{N\cdot s \cdot m^{-2}}$, respectively, whereas those of water outside the bubble are $\rho_{water}=1000\,\mathrm{kg\cdot m^{-3}}$ and $\mu_{water}=10\,\mathrm{N\cdot s \cdot m^{-2}}$, respectively, resulting in a density ratio of 1000 and a viscosity ratio of 100. The surface tension coefficient is $\sigma_\phi=1.96\, \mathrm{N\cdot m^{-1}}$, the gravitational acceleration is $\bm{g}=(0,-0.98)\, \mathrm{m\cdot s^{-2}}$, and the interface thickness and mobility are $\xi_\phi=0.01$ m and $M_\phi=10^{-4} \, \mathrm{m^4 \cdot N^{-1}\cdot s^{-1}}$, respectively. The Reynolds number, defined as the ratio of inertial to viscous effects, and the Eotvos number, which is the ratio of gravitation to surface-tension effects, are given by:
\begin{equation}
	\begin{aligned}
		Re =\frac{\rho_{water} U_0 D}{\mu_{water}}, \quad 
		Eo =\frac{\rho_{water} U_0^2 D}{\sigma_\phi},
	\end{aligned}
	\label{eq: Re_Eo}
\end{equation}
which are 35 and 125, respectively, where $D=2r_0$ is the characteristic length, and $U_0=\sqrt{2gr_0}$ is the characteristic velocity, with $g$ denoting the magnitude of the gravitational acceleration vector $\bm{g}$. To assess the simulation accuracy, three benchmark quantities, the center of mass, rising velocity, and circularity, are defined as follows:

\begin{equation}
	\begin{aligned}
		y_c =\frac{\int_{\Omega} y \frac{1+\phi}{2} d \Omega}{\int_{\Omega} \frac{1+\phi}{2} d \Omega}, \quad
		v_c =\frac{\int_{\Omega} v \frac{1+\phi}{2} d \Omega}{\int_{\Omega} \frac{1+\phi}{2} d \Omega}, \quad 
		\psi_c =\frac{P_a}{P_b}=\frac{2 \sqrt{\int_{\phi>0} \pi d \Omega}}{P_b},
	\end{aligned}
	\label{eq: three metrics}
\end{equation}
where $P_a$ is the equivalent perimeter of the circle having the same area as the bubble, and $P_b$ is the initial perimeter of the bubble. The center of mass $y_c$ and rising velocity $v_c$ quantify the dynamics of the bubble. The circularity $\psi_c$ quantifies the shape of the bubble: it equals 1 for a circular bubble and falls below 1 when deformation occurs.

In this case, the network comprises a Fourier embedding with 64 learnable frequencies, $N=3$ discontinuity-aware adaptive residual blocks (each containing 3 discontinuity-aware layers with 128 neurons), a learnable artificial-viscosity coefficient, and a linear output layer with 4 nodes corresponding to $u$, $v$, $p$, and $\phi$, respectively. It is trained for 20000 iterations using the AdamW optimizer under a cosine annealing learning rate schedule decaying from $10^{-3}$ to $10^{-5}$, on 2 Hygon G86 7285 with 32 GB of memory each.

\begin{figure}[!h]
	\centering
	\includegraphics{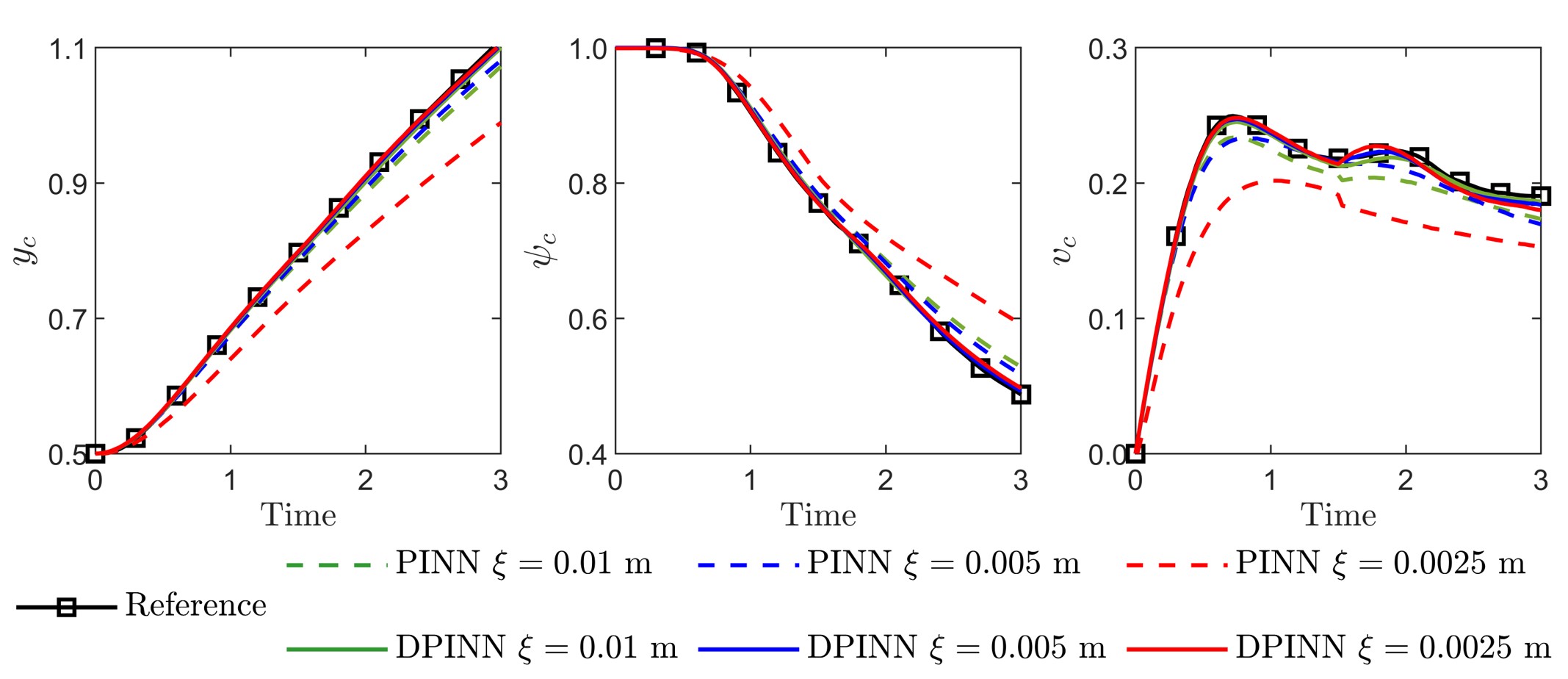}
	\caption{Comparison of three benchmark quantities for PINN and DPINN solutions at different $\xi_\phi$: (a) center of mass, (b) circularity, and (c) rising velocity.}
	\label{fig: bubble_statistics}
\end{figure}

During the evolution, the bubble first stretches horizontally, then deforms at its bottom part, and finally reaches a stable shape and a terminal rising velocity. Figure \ref{fig: bubble_conrourf} shows the temporal evolution of the phase field at $t=0$ s, $1.5$ s, and $3.0$ s for the conventional PINN and DPINN, where the top, middle, and bottom rows correspond to $\xi_\phi=0.01$ m, $0.005$ m, and $0.0025$ m, respectively. Figure \ref{fig: bubble_statistics} shows the temporal evolution of the three benchmark quantities, where the interfaces of all results are extracted from the $\phi=0$ contour line, and the reference solutions are from Ref.~\cite{aland2012benchmark}. For $\xi_\phi=0.01$ m, PINN captures the smooth interface evolution accurately. As $\xi_\phi$ decreases to 0.005 m, the interface becomes narrower, and the three benchmark quantities converge towards the reference solutions. However, as $\xi_\phi$ is further reduced to 0.0025 m, the interface becomes more dissipative instead, and the quantities deviate significantly from the reference due to the inherent limitation of the traditional network architectures in solving PDEs with sharp spatial transitions or rapid temporal variations. In contrast, DPINN consistently converges to the sharp interface solution, and the benchmark quantities show improved agreement with the reference solutions.

Table \ref{tab:error_norms_compare} summarizes the $R_1$, $R_2$, and $R_\infty$ error norms of the three benchmark quantities evaluated over the entire time horizon. DPINN consistently outperforms PINN across all norm measures and interface thicknesses $\xi_\phi$, particularly at $\xi_\phi=0.0025$ m, where the error of PINN increases significantly, whereas DPINN exhibits a further decrease in error, demonstrating its ability to resolve sharp-interface evolution.

\begin{table}[!h]
	\centering
	\renewcommand\arraystretch{0.8}
	\begin{tabular}{@{}lcccccccccc@{}}
		\toprule
		\multirow{2}{*}{Model} & \multirow{2}{*}{$\xi$/m} & \multicolumn{3}{c}{Center of mass} & \multicolumn{3}{c}{Circularity} & \multicolumn{3}{c}{Rising velocity} \\ \cmidrule(l){3-11}
		& & $R_1$ & $R_2$ & $R_{\infty}$ & $R_1$ & $R_2$ & $R_{\infty}$ & $R_1$ & $R_2$ & $R_{\infty}$ \\ \midrule
		\multirow{3}{*}{\centering PINN}
		& 0.0100 & 0.0211 & 0.0258 & 0.0356 & 0.0191 & 0.0259 & 0.0413 & 0.0640 & 0.0660 & 0.0871 \\
		& 0.0050 & 0.0155 & 0.0183 & 0.0274 & 0.0168 & 0.0206 & 0.0304 & 0.0468 & 0.0534 & 0.0841 \\
		& 0.0025 & 0.0761 & 0.0865 & 0.1101 & 0.0590 & 0.0722 & 0.1040 & 0.1999 & 0.2032 & 0.2597 \\ 
		\midrule
		\multirow{3}{*}{\centering DPINN} 
		& 0.0100 & 0.0059 & 0.0072 & 0.0105 & 0.0057 & 0.0067 & 0.0087 & 0.0165 & 0.0174 & 0.0226 \\
		& 0.0050 & 0.0027 & 0.0036 & 0.0069 & 0.0029 & 0.0034 & 0.0055 & 0.0131 & 0.0165 & 0.0266 \\
		& 0.0025 & 0.0025 & 0.0028 & 0.0053 & 0.0038 & 0.0056 & 0.0104 & 0.0169 & 0.0212 & 0.0420 \\ 
		\bottomrule
	\end{tabular}
	\caption{Error norms of the benchmark quantities (center of mass, circularity, and rising velocity) at different interface thickness parameters $\xi$ for PINN and DPINN.}
	\label{tab:error_norms_compare}
\end{table}

\subsection{Three-phase droplet icing}
\label{sub:3.3}
\begin{figure}[!h]
	\centering
	\includegraphics{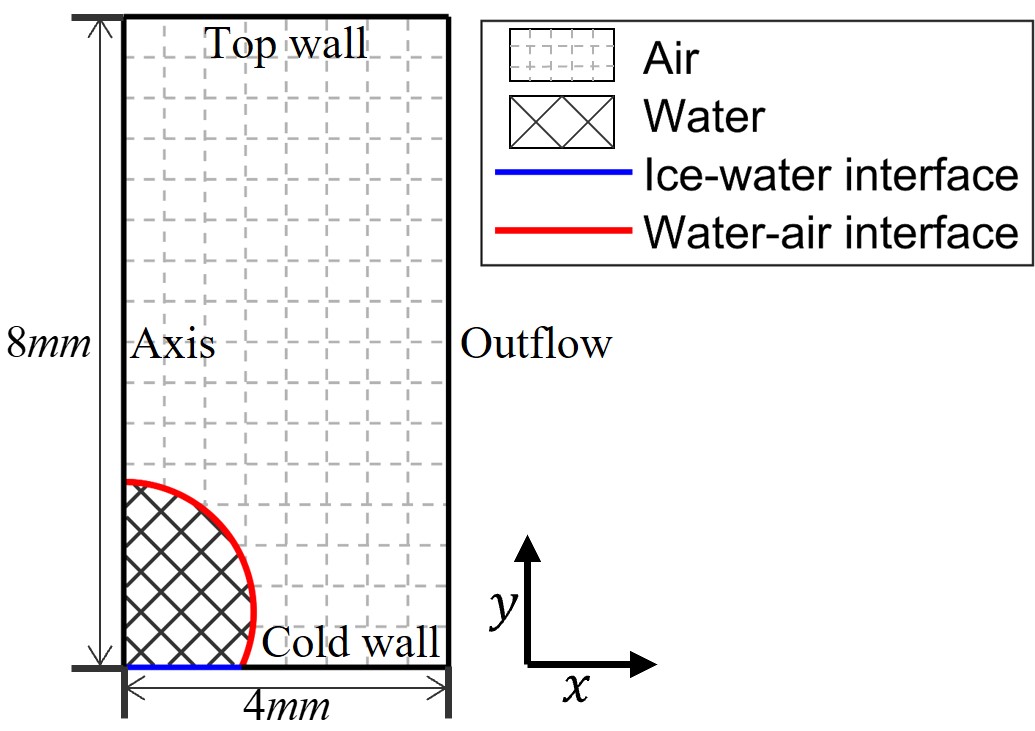}
	\caption{Initial and boundary conditions for the droplet icing case.}
	\label{fig: domain}
\end{figure}

The droplet icing process with phase change is fundamentally a moving-interface problem involving multiple interfaces: the ice-water interface, described by the AC equation (Eq. \ref{eq: AC}) modelling the freezing process, and the air–water interface, described by the CH equation (Eq. \ref{eq: CH}) capturing the volume-conserved evolution of the water–air boundary, with the flow field governed by the NS equations (Eqs. \ref{eq: mass}-\ref{eq: energy}). As shown in Figure \ref{fig: domain}, the spatial domain is defined as $\Omega=[0,4]$ mm $\times$ $[0,8]$ mm, discretized into 200 and 400 grid points in the $x$- and $y$-directions, respectively, and the temporal domain $\mathcal{T}=[0,12]$ s is divided into $M=3$ subdomains, $\mathcal{T}_1=[0,4]$ s, $\mathcal{T}_2=[4,8]$ s, and $\mathcal{T}_3=[8,12]$ s, each uniformly discretized with 200 time steps. The initial shape of the droplet is approximated by a circular segment with a radius of $r_0=1.6$ mm and a height of 2.28 mm, surrounded by air. To reduce computational cost, half of the droplet is simulated by applying a symmetry boundary condition along the central vertical axis. The bottom boundary is prescribed as a supercooled wall at $T_{sub}={-25^\circ\mathrm{C}}$ to initiate freezing, while the top boundary is maintained at a constant temperature. The initial temperature of both the water and air phases is set to ${25\,^\circ\mathrm{C}}$. The right boundary is specified as an outflow.

The material properties of pure ice, water, and air are listed in Table \ref{tab:physical_parameters} and are assumed to be constant and independent of temperature and pressure. In this case, the viscosity and density ratios exceed seven and three orders of magnitude, respectively, significantly exceeding previous cases and thereby increasing the challenge of simulations. In addition, the viscosity of ice is set to five orders of magnitude higher than that of water, which is sufficiently large to suppress any motion within the solid phase. The dynamics of the droplet icing process are governed by the dimensionless Prandtl number $Pr$, the Weber number $We$, and the Stefan number $St$:
\begin{equation}
	\begin{aligned}
		{Pr}=\frac{c_{p, water} \mu_{water}}{\kappa_{water}}, \;
		We=\frac{\rho_{water} U_0^2 r_0}{\sigma_c}, \;
		St=\frac{\left(T_M-T_{sub}\right) c_{p, water} \rho_{water}}{H}.
	\end{aligned}
	\label{eq: dimensionless}
\end{equation}
Their values are $Pr=7.1$, $We=1.7 \times 10^{-7}$, and $St=0.31$, respectively, with $U_0=r_0/\tau_0$ is the characteristic velocity, where ${\tau_0}={\rho_{water} c_{p,water}r_0^2}/{\kappa_{water}}$ is the characteristic time. 

\begin{table}[!h]
	\centering
	\small
	\renewcommand\arraystretch{0.85}
	\setlength{\tabcolsep}{5pt}
	\begin{tabular}{@{}lccc@{}}
		\toprule
		Parameter & Symbol & Value & Unit \\
		\midrule
		Density
		& $\rho_{air},\,\rho_{water},\,\rho_{ice}$
		& $1,\;1000,\;898$
		& $\mathrm{kg\cdot m^{-3}}$ \\
		
		Dynamic viscosity
		& $\mu_{air},\,\mu_{water},\,\mu_{ice}$
		& $1\times10^{-5},\;1\times10^{-3},\;1\times10^{2}$
		& $\mathrm{N\cdot s \cdot m^{-2}}$ \\
		
		Thermal conductivity
		& $\kappa_{air},\,\kappa_{water},\,\kappa_{ice}$
		& $0.0209,\;0.5918,\;2.25$
		& $\mathrm{W\cdot m^{-1}\cdot K^{-1}}$ \\
		
		Heat capacity
		& $c_{p,air},\,c_{p,water},\,c_{p,ice}$
		& $1003,\;4200,\;2018$
		& $\mathrm{J\cdot kg^{-1}\cdot K^{-1}}$ \\
				
		Surface tension coefficient
		& $\sigma_c,\,\sigma_\phi$
		& $3.17\times10^{-2},\;7.27\times10^{-2}$
		& $\mathrm{N\cdot m^{-1}}$ \\
		
		Interface thickness
		& $\xi_c,\,\xi_\phi$
		& $8.00\times10^{-5},\;8.00\times10^{-5}$
		& $\mathrm{m}$ \\

		Mobility of ice-water interface
		& $M_c$
		& $2.00\times10^{-3}$
		& $\mathrm{m^2 \cdot N^{-1}\cdot s^{-1}}$ \\
		
		Mobility of water-air interface
		& $M_\phi$
		& $2.50\times10^{-11}$
		& $\mathrm{m^4 \cdot N^{-1}\cdot s^{-1}}$ \\

		Latent heat per unit volume
		& $H$
		& $3.34\times10^{8}$
		& $\mathrm{J\cdot m^{-3}}$ \\
		
		Melting temperature
		& $T_M$
		& $2.73\times10^{2}$
		& $\mathrm{K}$ \\
		\bottomrule
	\end{tabular}
	\caption{Material properties and physical parameters used in the simulations.}
	\label{tab:physical_parameters}
\end{table}

The network comprises a Fourier embedding with 64 learnable frequencies, $N=3$ discontinuity-aware adaptive residual blocks (each containing 3 discontinuity-aware layers with 128 neurons), a learnable artificial-viscosity coefficient, and a linear output layer with 6 nodes corresponding to $u$, $v$, $p$, $T$, $c$, and $\phi$, respectively. It is trained for 20000 iterations using the AdamW optimizer under a cosine annealing learning rate schedule decaying from $5 \times 10^{-4}$ to $10^{-7}$, on 2 Hygon G86 7285 with 32 GB of memory each.

\begin{figure}[!h]
	\centering
	\includegraphics{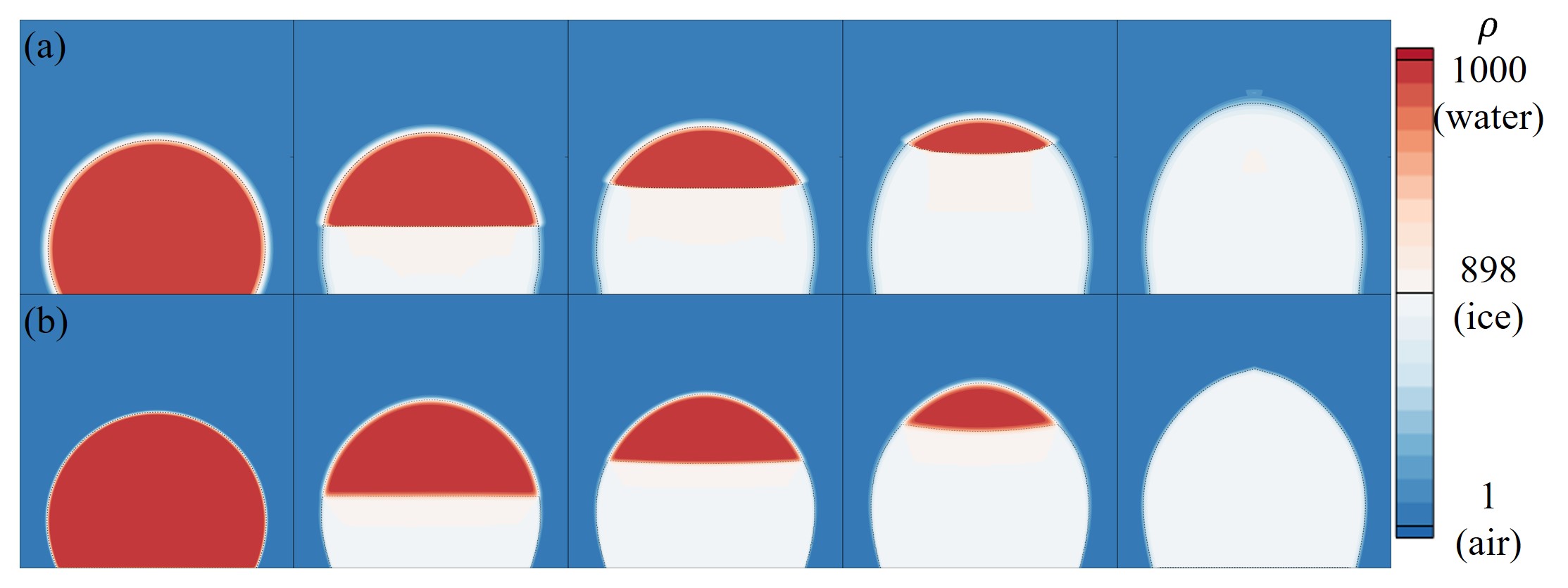}
	\caption{Comparison of the density field evolution during the droplet icing process: (a) reference numerical simulation using the same method as Ref.~\cite{wang2025energy}, and (b) DPINN. Dotted lines represent the interfaces, corresponding to $\phi=0$ and $c=-0.5$.}
	\label{fig: droplet_p1}
\end{figure}

Figure \ref{fig: droplet_p1}a and \ref{fig: droplet_p1}b present the evolution of the density field during the icing process obtained from the numerical simulation and DPINN, respectively. The numerical results are from Ref.~\cite{wang2025energy}, utilizing a $200\times 200\times 2$ grid in a Hele-Shaw cell with a time step of $1.6\times 10^{-6}$ s and $2.4\times10^7$ iterations executed in parallel on 50 CPU cores (Hygon G86 7285). The freezing of a water droplet on a cold surface is a typical heterogeneous nucleation process. During the early stage, solidification initiates at the base, while the droplet height increases gradually due to expansion. Subsequently, the premature formation of a thin ice shell due to surface cooling by the surrounding air suppresses lateral deformation and directs the expansion vertically, resulting in an increase in the droplet height while its width remains constant and the development of a concave freezing front with an increasing height difference \cite{wang2025energy}. As a result, this upward expansion eventually produces a tip similar to a cone. DPINN accurately reproduces this process and captures the sharp interfacial dynamics, yielding an icing shape featuring a point tip that agrees with published numerical results \cite{wang2025energy}.

\begin{figure}[!h]
	\centering
	\includegraphics{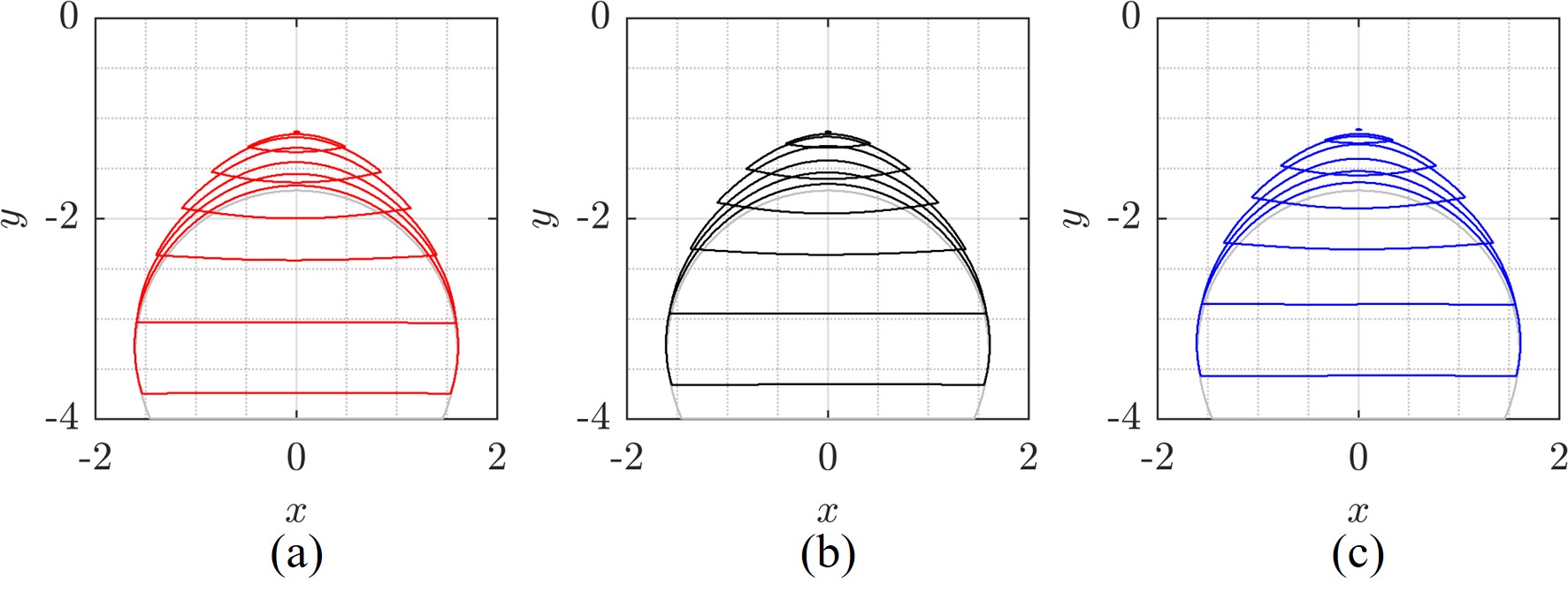}
	\caption{Droplet profiles and ice front evolutions at different substrate temperatures: 
		(a) $T_{\mathrm{sub}}=-20^\circ\mathrm{C}$, 
		(b) $T_{\mathrm{sub}}=-25^\circ\mathrm{C}$, 
		(c) $T_{\mathrm{sub}}=-30^\circ\mathrm{C}$. 
		From bottom to top, the nearly horizontal lines depict the evolution of the ice front, with a time interval of 2 s between adjacent lines.}
	\label{fig:ICE_T}
\end{figure}

In addition, we examine the influence of different substrate supercooling conditions on the freezing behavior. Figure \ref{fig:ICE_T}a-c show the evolution of the freezing front driven by substrate temperatures $T_{sub}$ of $-20^\circ\mathrm{C}$, $-25^\circ\mathrm{C}$, and $-30^\circ\mathrm{C}$, corresponding to Stefan numbers $St=0.25$, $0.31$, and $0.38$, respectively. The lower temperature of the substrate speeds up the freezing process. Due to the cooling effect of the air, the region of the droplet near the air reaches a lower temperature first, causing the freezing front to evolve from an approximately horizontal shape to a concave one, agreeing well with the experiments \cite{marin2014universality}.

\section{Conclusion}
\label{sub:4}
A discontinuity-aware physics-informed neural network, referred to as DPINN, is proposed to improve the capability of AI-based simulators in solving the coupled AC–CH–NS equations for three-phase flows with phase changes and sharp interfaces without relying on prior measured data. In terms of the network architecture, the method employs a Fourier embedding layer to mitigate spectral bias, learnable discontinuity-aware activation functions to automatically detect and model sharp interfaces, and a learnable local artificial viscosity term to stabilize the algorithm in regions of steep gradients. For optimization, time-marching and adaptive loss-weighting strategies are introduced to reduce long-horizon errors and mitigate gradient conflicts in the coupled phase-field formulation, respectively. This method is validated on the two-phase reversed single-vortex and bubble-rising problems, accurately resolving the sharp interface dynamics and significantly outperforming conventional PINNs. DPINN is further applied to a three-phase droplet-icing case with phase change, where the viscosity and density ratios between ice, water, and air exceed seven and three orders of magnitude, respectively, accurately reproducing the freezing process and the sharp pointy-tip formation.

It is worth noting that the training cost and generalizability of the proposed method can be further improved. In terms of training cost, the AC and NS equations with artificial viscosity involve second-order derivatives, while the CH equation even requires fourth-order derivatives, leading to high computational costs via automatic differentiation. Furthermore, the time-marching strategy, which approximates the solution manifold in a piecewise manner to reduce long-horizon errors by training multiple sequential networks, significantly increases the overall computational cost. In terms of generalizability, the high computational expense and the piecewise approximation strategy limit the development of a parameterized multiphase flow solver and its extension to more practical 3D scenarios. In future research, model reduction strategies, whether based on lightweight neural network architectures or the reduced-order methods for governing equations, and sequence modeling approaches, such as causal training and attention mechanisms to mitigate long-term error accumulation without the overhead of piecewise time-marching, will be introduced to extend the DPINN framework to large-scale parameterized problems and facilitate practical engineering applications.

\section*{Declaration of competing interest}
The authors declare that they have no known competing financial interests or personal relationships that could have appeared to influence the work reported in this paper.
\section*{Data availability}
Data will be made available on request.
\section*{Acknowledgments}
The authors would like to acknowledge the support from the National Natural Science Foundation of China (Grant Nos. 12202059).

\bibliographystyle{elsarticle-num-names} 
\bibliography{ref-ice}
\biboptions{numbers,sort&compress}






\end{document}